\documentclass[conference,twocolumn,romanappendices]{IEEEtran}
\usepackage{color}
\usepackage[left=1in, top=1in, right=1in, bottom=1in]{geometry}
\usepackage{setspace}
\usepackage{array}
\usepackage{cite}
\usepackage{multirow}
\usepackage{stmaryrd}
\usepackage{setspace}
\usepackage{amssymb}
\usepackage[cmex10]{amsmath}
\usepackage{amsthm}
\usepackage[pdftex]{graphicx}
\usepackage{epstopdf}
\usepackage{epsfig}
\usepackage{verbatim}
\usepackage{bbding}
\usepackage{dsfont}

\newcommand{\ma}[1]{\ensuremath\mathcal{#1}}
\newcommand{\bs}[1]{\ensuremath\boldsymbol{#1}}

\newsavebox{\subfigure}
\newsavebox{\theorembox}
\newsavebox{\lemmabox}
\newsavebox{\corollarybox}
\newsavebox{\propositionbox}
\newsavebox{\examplebox}
\newsavebox{\conjecturebox}
\newsavebox{\algbox}
\newsavebox{\qbox}
\newsavebox{\problembox}
\newsavebox{\definitionbox}
\newsavebox{\assumptionbox}
\newsavebox{\hypothesisbox}
\newsavebox{\factbox}
\newsavebox{\remarkbox}
\savebox{\theorembox}{\noindent\bf Theorem}
\savebox{\lemmabox}{\noindent\bf Lemma}
\savebox{\corollarybox}{\noindent\bf Corollary}
\savebox{\propositionbox}{\noindent\bf Proposition}
\savebox{\examplebox}{\noindent\bf Example}
\savebox{\conjecturebox}{\noindent\bf Conjecture}
\savebox{\algbox}{\noindent\bf Algorithm}
\savebox{\qbox}{\noindent\bf Question}
\savebox{\definitionbox}{\noindent\bf Definition}
\savebox{\problembox}{\noindent\bf Problem}
\savebox{\assumptionbox}{\noindent\bf Assumption}
\savebox{\hypothesisbox}{\noindent\bf Hypothesis}
\savebox{\factbox}{\noindent\bf Fact}
\savebox{\remarkbox}{\noindent\bf Remark}
\newtheorem{theorem}{\usebox{\theorembox}}
\begin{document}
\title{\huge Analysis and Design of Irregular Graphs for\\
 Node-Based Verification-Based Recovery Algorithms in\\
  Compressed Sensing}
\author{Yaser Eftekhari, Amir H. Banihashemi, Ioannis Lambadaris\\
			Department of Systems and Computer Engineering, Carleton University, Ottawa, Canada}
\maketitle
\thispagestyle{empty}
\begin{abstract}
In this paper, we present a probabilistic analysis of iterative node-based verification-based (NB-VB) recovery algorithms over irregular graphs in the context of compressed sensing. Verification-based algorithms are particularly interesting due to their low complexity (linear in the signal dimension $n$). The analysis predicts the average fraction of unverified signal elements at each iteration $\ell$ where the average is taken over the ensembles of input signals and sensing matrices. The analysis is asymptotic ($n \rightarrow \infty$) and is similar in nature to the well-known density evolution technique commonly used to analyze iterative decoding algorithms. Compared to the existing technique for the analysis of NB-VB algorithms, which is based on numerically solving a large system of coupled differential equations, the proposed method is much simpler and more accurate. This allows us to design irregular sensing graphs for such recovery algorithms. The designed irregular graphs outperform the corresponding regular graphs substantially. For example, for the same recovery complexity per iteration, we design irregular graphs that can recover up to about $40\%$ more non-zero signal elements compared to the regular graphs. Simulation results are also provided which demonstrate that the proposed asymptotic analysis matches the performance of recovery algorithms for large but finite values of $n$.
\end{abstract}
\section{Introduction}
\label{firstintro}
Consider a signal $\bs{v}\in\mathbb{R}^n$ with only $k$ non-zero elements. Let $m$ be a positive integer so that $k<m\ll n$. The main idea in compressed sensing is to represent $\bs{v}$ with measurements $\bs{c}\in\mathbb{R}^m$ (\textit{measuring process}), and to be able to recover back the original signal $\bs{v}$ from the measurements $\bs{c}$ (\textit{recovery process}) \cite{D06,CRTFeb06}. 

The measuring process is essentially a linear transformation that can be represented by the matrix multiplication $\bs{c} = \bs{G}\bs{v}$, or equivalently characterized by a bipartite graph \cite{XH07}. In these representations, $\bs{G}$ is referred to as the \emph{sensing matrix} and the bipartite graph as the \textit{sensing graph}, respectively. The recovery process is essentially estimating $\bs{v}$ based on the knowledge of $\bs{c}$ and $\bs{G}$ and is considered successful if $\bs{v}$ is estimated correctly. 

If the density of edges in the sensing graph is high, the computational complexity of finding the measurements and that of the recovery is considerably higher compared to the cases where the sensing graph is sparse. The high complexity of recovery for dense graphs hinders their application to high-dimensional signal recovery (signals with large $n$). Moreover, in certain compressed sensing applications such as computer networks \cite{AJWAK10, LMP08, LMPDK08}, channel coding \cite{CT05}, spectrum sensing \cite{MYLHH10}, and identification of linear operators \cite{HB11}, the nature of the problem results in a formulation with a sparse sensing graph. The main focus of this paper is on such problems. More specifically, our main interest is in a sub-class of message-passing recovery algorithms, called \textit{Node-Based Verification-Based} (NB-VB) algorithms \cite{ZP07J,ZP09,EHBL11,EHBLISIT11}, and the performance of these algorithms over {\em irregular} sparse graphs (matrices). The VB recovery algorithms, in general, are iterative and have computational complexity $\ma{O}(n)$, which makes them suitable for applications involving recovery of signals with large $n$. Moreover, if certain conditions are satisfied, the performance of VB algorithms is not sensitive to the distribution of non-zero elements in the sensing matrix and the original signal \cite{EHBLISIT11, EHBL11}. Another interesting feature of VB algorithms is that their performance can be analyzed in the asymptotic case ($n \rightarrow \infty$) \cite{LM05,ZP07J,ZP07,ZP09,EHBL11,EHBLISIT11}. These properties make the VB algorithms a suitable choice for low-complexity recovery of sparse signals. For a comprehensive study on the VB algorithms, we refer the interested readers to \cite{BGIKS08, EHBL11}. 

The VB algorithms are, however, sensitive to the presence of noise in the measured data. The authors in \cite{EHBL11} discussed the use of a thresholding technique to deal with noisy measurements. This technique is very effective in the high signal to noise ratio (SNR) regime (such as the scenario in \cite{AJWAK10}). Having said that, some compressed sensing applications (see \cite{LMPDK08,LMP08} for some examples) are formulated as noiseless problems with sparse measurement matrices. Furthermore, the noise-free analysis of recovery algorithms serves as an upper bound for the performance of the noisy versions. All of the above highlight the importance of noise-free analysis of the recovery algorithms in compressed sensing, in general, and in compressed sensing with sparse measurement matrices, in particular.

The main focus of this paper is to analyze NB-VB recovery algorithms for compressed sensing problems with irregular sensing graphs and to design sensing graphs that perform well with these recovery algorithms. Our results are derived in the asymptotic regime ($n \rightarrow \infty$). In this regime, we assume a probabilistic model for the input signal, in which a signal element is zero with probability $1- \alpha$ or takes a value from a continuous distribution with probability $\alpha$. Henceforth, the parameter $\alpha$ is referred to as the \emph{density factor}. Let $\alpha^{(\ell)}$ denote the probability that a signal element is non-zero and unknown at iteration $\ell$ over the ensemble of all sensing graphs and all inputs of interest. In the asymptotic regime, the recovery algorithm is called successful for the initial density factor $\alpha^{(0)} = \alpha$ if and only if $\lim_{\ell \rightarrow \infty} \alpha^{(\ell)} = 0$. Indeed, if the initial density factor is smaller than a certain threshold, referred to as the \textit{success threshold}, then the recovery algorithm is successful as $n \rightarrow \infty$ and $\ell \rightarrow \infty$ \cite{LM05,ZP07J,ZP07,ZP09,EHBL11,EHBLISIT11}. Fixing the compression ratio $m/n$, it is desirable to devise sensing graphs with the highest success threshold possible.

An asymptotic analysis of NB-VB algorithms was first presented in \cite{ZP07J}, where a system of coupled differential equations had to be solved. To cope with the high complexity of solving such a systems, the authors used numerical methods to solve the system for finite values of $n$, and thus obtained an approximation of the asymptotic result by choosing a large value of $n$. The numerical approximation in \cite{ZP07J} translates to long running times and the possibility of numerical errors propagating through the iterations in the analysis. The latter would compromise the accuracy of the obtained success thresholds \cite{EHBLISIT11,EHBL11}.

In \cite{EHBLISIT11,EHBL11}, the authors developed a low-complexity framework for the asymptotic analysis of NB-VB algorithms over sparse random {\em regular} sensing graphs. The analysis presented in \cite{EHBLISIT11,EHBL11} was significantly faster and more robust against numerical errors compared to the approach of \cite{ZP07J}.

In this paper, we extend the analysis presented in \cite{EHBLISIT11,EHBL11} to irregular graphs. Our simulations show that for a given compression ratio $m/n$, irregular graphs can provide up to $40\%$ larger success thresholds compared to regular graphs. In this comparison, since the number of edges in both graphs is the same, the recovery complexity remains almost the same. Just like the analysis in \cite{EHBLISIT11,EHBL11}, the proposed analysis is developed for noiseless measurements and its computational complexity increases only linearly with the number of iterations. Moreover, the analysis is simple to perform, requiring only additions and multiplications. This makes it possible to use the analysis at the core of a process to design degree distributions for irregular sensing graphs that perform well with NB-VB algorithms. The performance measure considered in this work is the success threshold. 
\section{Ensembles of Sensing Graphs and Inputs}
\label{back_knowledge}
Let $\ma{G}({V}\cup{C},{E})$ denote a bipartite graph or a bigraph with the node set ${V}\cup{C}$ and the edge set ${E}$, so that every edge in ${E}$ connects a node in ${V}$ to a node in ${C}$. Further, let $\bs{A}(\ma{G})$ denote the \textit{biadjacency matrix} of graph $\ma{G}$; the entry $a_{ij}$ in $\bs{A}$ is $1$ if there exists an edge connecting the nodes $c_i\in C$ and $v_j\in V$. Following the coding terminology, we refer to the sets ${V}$ and $C$ as \textit{variable nodes} and \textit{check nodes}, respectively.

In general, a bigraph can be \textit{weighted} and \textit{irregular}. In the weighted bigraph $\ma{G'}({V}\cup{C},W(E))$ a weight $w_{ij}:= w(e_{ij})\in \mathbb{R} \backslash \{0\}$ is associated with each edge $e_{ij}\in E$. The weight $w_{ij}$ also appears in the $(i,j)$th entry of the biadjacency matrix $\bs{A}(\ma{G'})$ corresponding to the weighted bigraph $\ma{G'}$.

In a bigraph, a node in $V$ ($C$) has degree $i$ if it is neighbor (connected) to $i$ nodes in $C$ ($V$). Let $\lambda_i \in \mathbb{R}^+$ and $\rho_i \in \mathbb{R}^+$ denote the fraction of nodes in $V$ and $C$ with degree $i$, respectively. The polynomials $\lambda(x) = \sum_i \lambda_i x^i$ and $\rho(x) = \sum_i \rho_i x^i$ are referred to as \emph{degree distributions} corresponding to nodes in $V$ and $C$, respectively. Clearly, $\lambda(1) = \rho(1) = 1$. For mathematical convenience, we define $\bar{d_v} := \sum_i i \lambda_i$ and $\bar{d_c} := \sum_j j \rho_j$ and we refer to them as the \emph{average variable degree} and the \emph{average check degree}, respectively.

For given $\lambda(x),\rho(x)$ and $n$, let $\ma{G}^{n}(\lambda(x),\rho(x))$ ($\ma{G}^{n}(\lambda,\rho)$ for short) denote the ensemble of all irregular bigraphs with $n$ variable nodes and degree distributions $\lambda(x)$ and $\rho(x)$. Further, let $\ma{W}_{f}^{m\times n}$ be the ensemble of all $m \times n$ ($m = n\bar{d_v}/\bar{d_c}$) matrices with i.i.d. entries $w$ drawn according to a distribution $f(w)$. Now, for any irregular bigraph $\ma{G}({V}\cup{C},E) \in \ma{G}^{n}(\lambda,\rho)$ and any weight matrix $\bs{W} \in \ma{W}_{f}^{m\times n}$, we form the corresponding $(n,\lambda,\rho)$-weighted irregular bigraph $\ma{G'}({V}\cup{C},W(E))$ as follows. Let us assume an arbitrary, but fixed, labeling scheme for node sets $V$ and $C$ over the ensemble $\ma{G}$. To every edge $e_{ij}\in E$, $1\leq i \leq m, 1 \leq j \leq n$, connecting $c_i \in C$ and $v_j \in V$, we assign the weight in row $i$ and column $j$ of the weight matrix $\bs{W}$; i.e., $w(e_{ij}) = w_{ij}$. Thus, we construct the ensemble of all $(n,\lambda,\rho)$-weighted irregular bigraphs, denoted by $\ma{G}_{f}^{n}(\lambda,\rho)$, by freely combining elements in $\ma{G}^{n}(\lambda,\rho)$ and $\ma{W}_{f}^{m\times n}$.

To describe the inputs of interest, let $\alpha\in [0,1]$ be a fixed real number and $\bs{v}$ be a vector of length $n$ with elements $\bs{v}_i$ drawn i.i.d. according to $\Pr[\bs{v}_i=v]=\alpha g(v)+(1-\alpha)\delta(v)$, where $\delta(\cdot)$ and $g(\cdot)$ are the Kronecker delta and a probability density function, respectively. We denote the ensemble of all such vectors by $\ma{V}_{g}^{n}(\alpha)$.\footnote{It is worth noting that the expected fraction of non-zero elements in such a vector is $\alpha$. Using a Chernoff bound, it can be shown that the actual fraction of non-zero elements in a randomly chosen vector from this ensemble is tightly concentrated around its expected value ($\alpha$) with high probability.}

To build the compressed sensing setup, let $\ma{G}(V\cup C,W(E))$ be a weighted irregular bigraph drawn uniformly at random from the ensemble $\ma{G}_f^{n}(\lambda,\rho)$ with $\bs{G}$ as its biadjacency matrix. Moreover, let $\bs{v}$ be a signal vector drawn uniformly at random from the ensemble $\ma{V}_{g}^{n}(\alpha)$. Also, let $\bs{c} = \bs{G} \bs{v}$. The sets of signal elements $\bs{v}$ and measurements $\bs{c}$ are respectively mapped to the vertex sets $V$ and $C$ ($|V|=n$, $|C|=m$) with the sensing matrix being the biadjacency matrix $\bs{G}$. Henceforth, we refer to the graph $\ma{G}$ as the \textit{sensing graph}. 
\section{VB Algorithms and Verification Rules}
\label{enc}
As a class of iterative recovery algorithms, the VB algorithms find the value of a set of signal elements in each iteration based on the knowledge of the measurements, the sensing graph and the previously verified signal elements. In \cite{SBB206,ZP08,ZP09,EHBL11,EHBLISIT11}, it was demonstrated that the continuity of at least one of the distributions $f$ or $g$ (non-zero weights of the sensing graph or non-zero signal elements), is a sufficient condition for the assigned value to a variable node at a certain iteration to be its true value with probability one (zero probability of false verification). In this paper also, we assume that the probability of false verification throughout the recovery algorithm is zero.

The algorithm discussed in \cite{SBB206}, which is the same as the LM2 algorithm of \cite{ZP09}, performs the best among all known VB algorithms in the context of compressed sensing \cite{EHBLISIT11,EHBL11,ZP07J}. We refer to this algorithm as SBB. The analysis of SBB is the focus in this work. The proposed analytical framework is, however, general and applicable to the analysis of other NB-VB algorithms. In what follows, we discuss the verification rules of the SBB algorithm.

When a variable node is verified at an iteration of an NB-VB algorithm, its value is subtracted from the value of all its neighboring check nodes. Then, the variable node and all its adjacent edges are removed from the sensing bigraph. Consequently, all the check nodes neighbor to the verified variable node face a reduction in their degree by one. The algorithm stops at an iteration if the set of verified variable nodes remains unchanged for two consecutive iterations or when all the variables are verified. A variable node in SBB is verified based on the following verification rules.

\begin{itemize}
	\item Zero Check Node (ZCN): All variable nodes neighbor to a zero-valued check node are verified with a value equal to zero.
	\item Degree One Check Node (D1CN): If a variable node is neighbor to a check node of degree 1, it is verified with the value of the check node.
	\item Equal Check Nodes (ECN): Let $\ma{C}$ denote a set of check nodes with the same non-zero value. Also, let $\ma{V}$ denote the set of all variable nodes neighbor to at least one check node in $\ma{C}$. Then, 1) a variable node in $\ma{V}$ which is not connected to all check nodes in $\ma{C}$ is verified with a value equal to zero; 2) if there exists a unique variable node in $\ma{V}$ that is neighbor to all check nodes in $\ma{C}$, then it is verified with the common value of the check nodes.
\end{itemize}

In order to introduce the asymptotic analysis of NB-VB algorithms and prove the concentration results, the authors in \cite{EHBL11,EHBLISIT11} developed a message passing representation of VB algorithms. In this paper, we use the same representation and refer the interested reader to the aforementioned references for more details.
\section{Asymptotic Analysis Framework}
\label{analysis}
Let the probability distributions $f$ and $g$, the degree distributions $\lambda$ and $\rho$, and the density factor $\alpha$ be fixed. It can be shown that the fraction of unverified non-zero variable nodes at each iteration $\ell$ of the SBB algorithm ($\alpha^{(\ell)}$) over a realization of the sensing graph $\ma{G} \in \ma{G}^{n}_f(\lambda,\rho)$ and a realization of the input signal $\ma{V} \in \ma{V}^{n}_g(\alpha)$ concentrates around the average of $\alpha^{(\ell)}$ taken over all the elements in the ensemble $\ma{G}^{n}_f(\lambda,\rho)\times \mathcal{V}^{n}_g(\alpha)$, as $n$ tends to infinity.\footnote{These {\em concentration results} have been proved in detail for regular graphs in \cite{EHBL11}. Similar results apply to the irregular sensing graphs with minor changes, and are therefore not presented in this paper.} The deterministic analysis presented here is to track the evolution of this average as $n$ goes to infinity. In the language of coding, the analysis is similar to the density evolution analysis of iterative decoding algorithms over irregular LDPC code ensembles, with the main difference being that the NB-VB algorithms do not conform to the principle of {\em extrinsic} message passing which significantly simplifies the density evolution analysis in the context of coding.

The mathematical framework for the analysis is similar to the one used in \cite{EHBLISIT11,EHBL11}, however with two extra variables $d_v$ and $d_c$ which represent the degree of a variable and a check node, respectively. These variables take different values for irregular graphs while for a regular graph they each have a fixed value. This makes the derivations more tedious. Due to the lack of space, in the following, we only provide a sketch of the analysis.

At the beginning of each iteration $\ell$, the analysis partitions the set of all unverified variable nodes into two (disjoint) sets: non-zero variable nodes ($\ma{K}^{(\ell)}$), zero-valued variable nodes ($\Delta^{(\ell)}$). Should the fraction of variable nodes in the set $\ma{K}^{(\ell)}$ tend to zero as iterations proceed, the fraction of variable nodes in the set $\Delta^{(\ell)}$ will also tend to zero and consequently the analysis declares a successful recovery \cite{EHBL11}.

Each iteration in the SBB algorithm is divided into two rounds ($R$), each consisting of two half-rounds ($HR$). In the first and second rounds, verified variable nodes belong to the sets $\ma{K}^{(\ell)}$ and $\Delta^{(\ell)}$, respectively. The configuration of the sets at the end of each half-round is specified using the superscript $(\ell,Rx,y)$, where $\ell$, $x \in \{1,2\}$ and $y \in \{1,2\}$ denote the iteration, round and half-round numbers, respectively.

We partition the set of all check nodes with the same degree (say $d_c$) into sets $\ma{N}^{(\ell)}_{i,j}(d_c), 0\leq i\leq d_c,0\leq j\leq d_c-i$, where $i$ and $j$ indicate the number of neighboring variable nodes in the sets $\ma{K}^{(\ell)}$ and $\Delta^{(\ell)}$, respectively. 

Let $\ma{K}^{(\ell)}(d_v)$ and $\Delta^{(\ell)}(d_v)$ denote the set of all non-zero and zero-valued unverified variable nodes with the same degree $d_v$, respectively. Then, the set $\ma{K}^{(\ell)}(d_v)$ is further divided into subsets $\ma{K}^{(\ell)}_{i}(d_v), 0\leq i\leq d_v,$ where $i$ denotes the number of neighboring check nodes in the set $\ma{N}^{(\ell)}_1 := \bigcup_{d_c}\bigcup_{j=0}^{d_c-1} \ma{N}^{(\ell,R1,1)}_{1,j}(d_c)$. Also, we divide the set $\Delta^{(\ell)}(d_v)$ into subsets $\Delta^{(\ell)}_i(d_v), 0\leq i\leq d_v,$ with the following definition: a variable node in $\Delta^{(\ell)}_i$ has $i$ neighboring check nodes which became zero-valued after HR1 of R2. Table \ref{T:changes} summarizes the sets affected in each half-round of each round at any iteration.

\begin{table}
\caption{Sets that change in each half-round of each round at any iteration}
\label{T:changes}
\begin{center}
\vspace{-.5 cm}
\renewcommand{\arraystretch}{1.5}
\begin{tabular}{|c|c|c|c|}
	\hline
	\multicolumn{2}{|c|}{R1} & \multicolumn{2}{c|}{R2} \\
	\hline
	HR1 & HR2 & HR1 & HR2 \\
	\hline
	$\ma{N}_{k,i} \rightarrow \ma{N}_{k,j}$ &	$\ma{K}_{i} \rightarrow \ma{K}_{j}$ & $\ma{N}_{i,k} \rightarrow \ma{N}_{j,k}$ & $\Delta_{i} \rightarrow \Delta_{j}$ \\
	\hline
\end{tabular}
\end{center}
\end{table}

Theorems \ref{SBBModel} and \ref{LMSBBModel} below, characterize the verification of unverified non-zero ($\ma{K}^{(\ell)}$) and zero-valued ($\Delta^{(\ell)}$) variable nodes at HR2-R1 and HR2-R2 in each iteration $\ell$ of the SBB algorithm, respectively. The proofs of the theorems are very similar to the ones presented in \cite{EHBL11} and therefore omitted.

\begin{theorem}
\label{SBBModel}
In the first round of any iteration $\ell$, a non-zero variable node of degree $d_v$ is verified if and only if it belongs to the set $\bigcup_{i=2}^{d_v}\ma{K}^{(\ell,R1,2)}_{i} \cup \hat{\ma{K}}^{(\ell,R1,2)}_{1}$, where the set $\hat{\ma{K}}^{(\ell,R1,2)}_{1}$ consists of all variable nodes in the set $\ma{K}^{(\ell,R1,2)}_{1}$ connected to the set $\ma{N}^{(\ell,R1,1)}_{1,0}$.
\end{theorem}

\begin{theorem}
\label{LMSBBModel}
In the second round of any iteration $\ell$, a zero-valued variable node of degree $d_v$ is verified if and only if it belongs to the set $\bigcup_{i=1}^{d_v}\Delta^{(\ell)}_{i}$.
\end{theorem}

The sets $\ma{K}^{(\ell)}$, $\Delta^{(\ell)}$, $\ma{N}^{(\ell-1,R2,1)}_{i,j}$, $\ma{K}^{(\ell-1,R1,2)}_{i}$, and $\Delta^{(\ell-1,R2,2)}_{i}$ fully describe the state of the algorithm at the beginning of iteration $\ell$. The probability that a variable node belongs to the set $\ma{K}^{(\ell)}$ is $\alpha^{(\ell)}$.

The asymptotic analysis tracks the probability that a node (variable node or check node) belongs to a certain set at each half-round, round, or iteration. The recovery is successful if and only if the probability $\alpha^{(\ell)}$ tends to zero, as $\ell$ tends to infinity. The analysis is based on the derivation of recursive equations that relate the probabilities described above for two consecutive iterations. The complexity of the analysis thus scales linearly with the number of iterations.
\section{Simulation Results}
\label{simulation}
To verify the asymptotic results obtained based on the analysis of Section \ref{analysis}, we perform some finite-length simulations for large values of $n$. The input signal in all simulations follows the probabilistic model described in Section \ref{back_knowledge}. Also, each non-zero signal element is drawn from a standard Gaussian distribution (zero-mean with variance one). The graphs are constructed randomly with no parallel edges and all edge weights are chosen to be $1$. Each simulation point is generated by averaging over 1000 random instances of the input signal. In simulations, the recovery is successful if and only if the input signal is recovered perfectly. For the analytical results, based on the fact that $\alpha^{(\ell)}$ is a non-increasing function of iteration number $\ell$, we consider the following stopping criteria:
\begin{enumerate}
	\item{Success:} $\alpha^{(\ell)} \leq 10^{-7}$.
	\item{Failure:} $\alpha^{(\ell)} > 10^{-7}$ and $|\alpha^{(\ell)} - \alpha^{(\ell-1)}| < 10^{-8}$.
\end{enumerate}

To calculate the success threshold, a binary search is performed within a certain range of initial density factors which includes the threshold. The search continues until the search region is smaller than $10^{-5}$.

To motivate the use of irregular graphs for the purpose of sparse signal recovery, we first present some comparison with regular graphs. In Table \ref{IrregularComp}, we compare the success threshold of the SBB algorithm over regular, left-regular (all variable nodes have the same degree), right-regular (all check nodes have the same degree) and bi-irregular graphs, when the average variable and check degrees are fixed at $\bar{d_v} = 4, \bar{d_c}=5$, respectively. The success thresholds reported for the left- and right-regular graphs are the highest thresholds obtained by optimizing right and left degree distributions with maximum degree $20$ and with up to four non-zero components, respectively. For the bi-irregular case, however, we restricted the search to only two degrees (bimodal distribution) both less than $20$ for both variable and check nodes. The optimized degree distributions $(\lambda(x),\rho(x))$ are: $(0.9310x^3 + 0.0350x^{17} + 0.0340x^{18},x^5)$, $(x^4,0.7100x^3 + 0.1830x^5 + 0.1070x^{20})$, and $(0.9000x^3+0.1000x^{13},0.9375x^4 + 0.0625x^{20})$ for right-regular, left-regular, and bi-irregular graphs, respectively. As expected, the bi-irregular graphs achieve the highest success threshold, almost $37\%$ higher than that of the regular graphs.

\renewcommand{\arraystretch}{1.5}
\begin{table}%
\centering
\caption{Success Thresholds for Regular, Left-Regular, Right-Regular, and Bi-Irregular Graphs with $\bar{d_v} = 4$, $\bar{d_c} = 5$}
\begin{tabular}{|c|c|c|}
\hline
Graph Type & Success Threshold & Improvement (\%) \\
\hline
\hline
Regular & $0.4225$ & - \\
\hline
Right-Regular & $0.5319$ & $25.89$ \\
\hline
Left-Regular & $0.5247$ & $24.16$ \\
\hline
Bi-Irregular & $0.5795$ & $37.15$ \\
\hline
\end{tabular}
\label{IrregularComp}
\end{table}

\begin{figure}[!h]
\centering
\includegraphics[width=3.5 in]{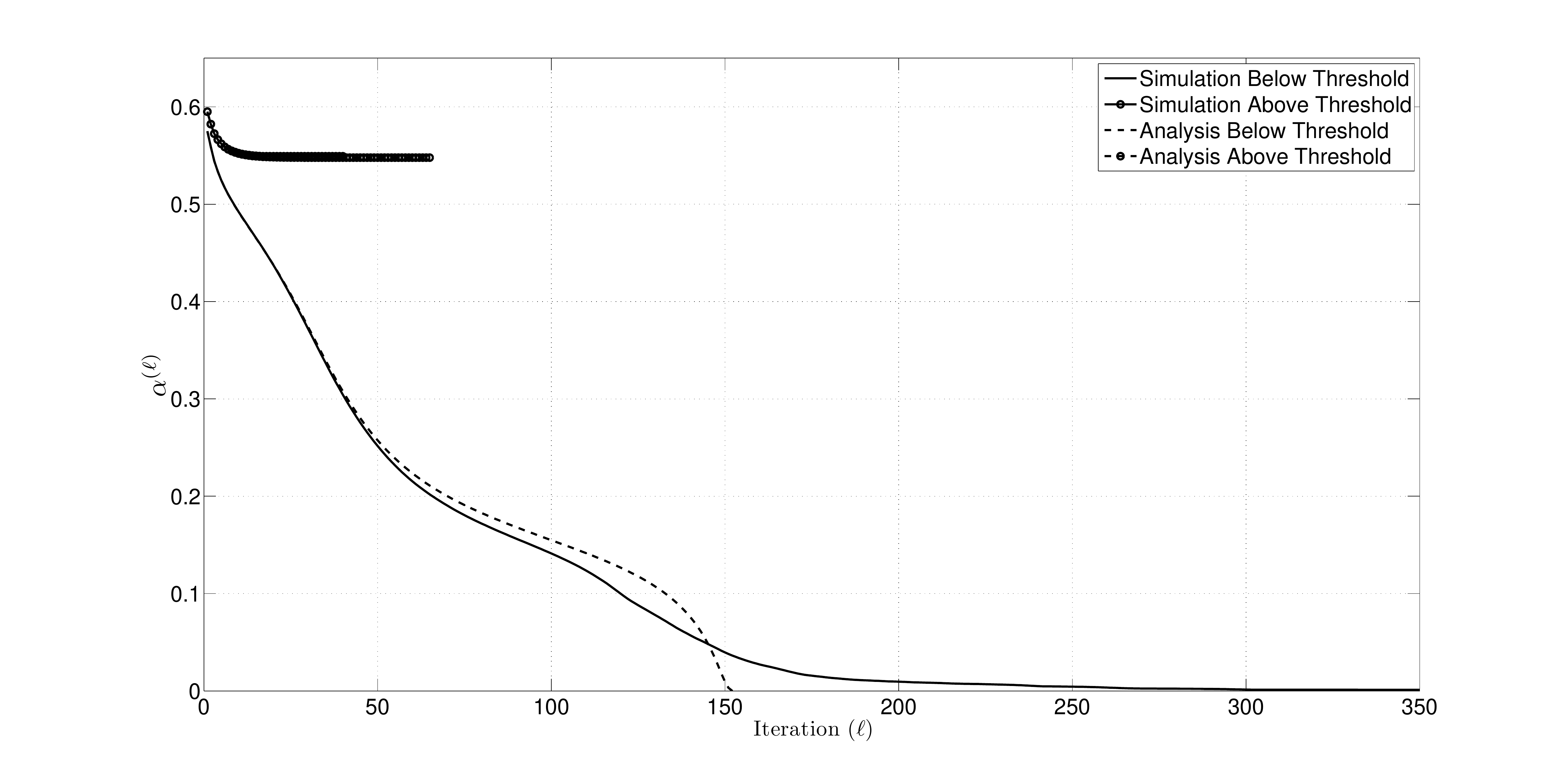}
\caption{Evolution of $\alpha^{(\ell)}$, obtained by the theoretical analysis, vs. iteration number $\ell$ (dotted line) and that of the normalized average unverified non-zero variable nodes vs. $\ell$ for $n=10^5$ (solid line).}
\label{Evolution100k}
\end{figure}

To investigate the degree of agreement between our asymptotic analysis and finite-length simulations, we have presented in Fig. \ref{Evolution100k} the evolution of $\alpha^{(\ell)}$ (for the theoretical results) and the average unverified non-zero variable nodes normalized by $n$ (for the finite-length results) with iterations $\ell$ for the SBB algorithm. The sensing graph is a randomly constructed bi-irregular graph with the optimized degree distributions $\lambda(x) = 0.9000 x^3 + 0.1000 x^{13}$ and $\rho(x)=0.9375 x^4 + 0.0625 x^{20}$ with $10^5$ variable nodes. Two values of $\alpha^{(0)}$ are selected: one above the success threshold ($0.595 > 0.5795$) and one below it ($0.575 < 0.5795$). The theoretical results are shown by dotted lines while simulations for $n = 10^5$ are presented with solid lines. As one can see, the two sets of results are in close agreement particularly for the cases where $\alpha^{(0)}$ is above the threshold and also for smaller values of $\ell$.

\begin{figure}[!h]
\centering
\includegraphics[width=3.5 in]{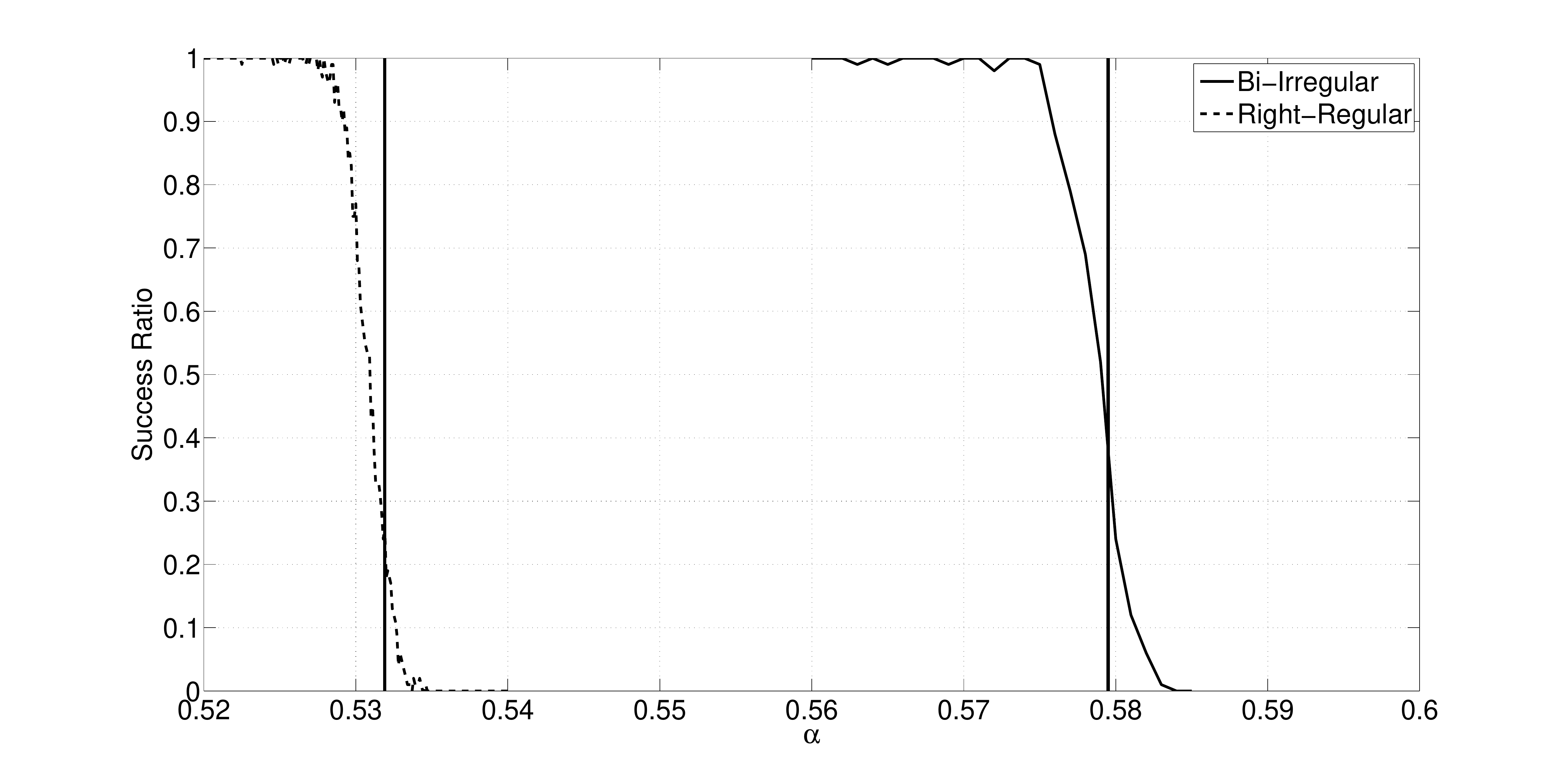}
\caption{Verifying the success threshold of SBB over right-regular and bi-irregular graphs of Table \ref{IrregularComp} through finite-length simulations.}
\label{Success}
\end{figure}

In Figure \ref{Success}, we have plotted the average normalized number of unverified variables (success ratio) of SBB over the graphs with optimized right-regular and bi-irregular degree distributions of Table \ref{IrregularComp} vs. the initial density factor. Each graph has $n=10^5$ variable nodes. We note that the success threshold for each graph, demonstrated by a vertical line on the figure, matches the waterfall region of the corresponding finite-length simulations.

More optimization results for right-regular graphs with $\bar{d_v} =4$ and different $d_c$ values are presented in Table \ref{Ex_Search}. For comparison, the threshold values $\alpha^*_R$ of the corresponding regular graphs are also given in the table. An inspection of the threshold values reveals an improvement of about $18\%$ to $26\%$ in the threshold of irregular graphs in comparison with that of regular graphs.

\begin{table}%
\centering
\caption{Optimal Left Degree Distributions and the Corresponding Thresholds $\alpha^*$ of SBB for Right-Regular Graphs with $\bar{d_v} = 4$ and Different Check Degrees Along with the Threshold $\alpha^*_R$ of the Corresponding Regular Graphs}
\begin{tabular}{|c|c|c|c|}
\hline
$d_c$ & Optimum $\lambda(x)$ & $\alpha^*$ & $\alpha_R^*$\\
\hline
\hline
5 & $0.931x^3 + 0.035x^{17} + 0.034x^{18}$ & $0.5319$ & $0.4225$\\
\hline
6 & $0.917x^3 + 0.082x^{15} + 0.001x^{19}$ & $0.4137$ & $0.3387$\\
\hline
7 & $0.906x^3 + 0.034x^{13} + 0.06x^{14}$ & $0.3369$ & $0.2811$\\
\hline
8 & $0.896x^3 + 0.04x^{12} + 0.064x^{13}$ & $0.2831$ & $0.2394$\\
\hline
\end{tabular}
\label{Ex_Search}
\end{table}

To reduce the design complexity of irregular graphs, we also investigate the effect of reducing the number of non-zero components in the left degree distribution of right-regular graphs. The results for some optimal bimodal left degrees are presented in Table \ref{BiModal}. It can be seen from the table that the difference between the thresholds $\alpha^*_B$ of bimodal left degrees and the thresholds $\alpha^*$ of left degrees with up to 4 nonzero components are negligible, i.e., only two different degrees on the variable side of a right-regular sensing graph  suffices to provide very good performance. 

\renewcommand{\arraystretch}{1.5}
\begin{table}%
\centering
\caption{Success Thresholds $\alpha^*_B$ of SBB over Right-Regular Graphs with Optimal Bimodal Left Degree Distributions Along with the Bimodal Degree Distributions ($\alpha^*$ is the SBB Threshold of a Right-Regular Graph with Optimal Left Degree Distribution with Up to 4 non-zero Components)}
\begin{tabular}{|c|c|c|c|}
\hline
$(\bar{d_v},d_c)$ & $\lambda(x)$ Bi-Modal & $\alpha^*_B$ & $\alpha^*$ \\
\hline
\hline
$(4,5)$ & $0.9333 x^3 + 0.0667 x^{18}$ & $0.5318$ & $0.5319$\\
\hline
$(4,6)$ & $0.9166 x^3 + 0.0834 x^{15}$ & $0.4137$ & $0.4137$\\
\hline
$(5,9)$ & $0.8750 x^3 + 0.1250 x^{19}$ & $0.3429$ & $0.3431$\\
\hline
$(5,10)$ & $0.8571 x^3 + 0.1429 x^{17}$ & $0.3001$ & $0.3017$\\
\hline
\end{tabular}
\label{BiModal}
\end{table}
\bibliographystyle{IEEEtran}
\bibliography{ISIT2012_biblio}

\begin{thebibliography}{10}
\providecommand{\url}[1]{#1}
\csname url@samestyle\endcsname
\providecommand{\newblock}{\relax}
\providecommand{\bibinfo}[2]{#2}
\providecommand{\BIBentrySTDinterwordspacing}{\spaceskip=0pt\relax}
\providecommand{\BIBentryALTinterwordstretchfactor}{4}
\providecommand{\BIBentryALTinterwordspacing}{\spaceskip=\fontdimen2\font plus
\BIBentryALTinterwordstretchfactor\fontdimen3\font minus
  \fontdimen4\font\relax}
\providecommand{\BIBforeignlanguage}[2]{{%
\expandafter\ifx\csname l@#1\endcsname\relax
\typeout{** WARNING: IEEEtran.bst: No hyphenation pattern has been}%
\typeout{** loaded for the language `#1'. Using the pattern for}%
\typeout{** the default language instead.}%
\else
\language=\csname l@#1\endcsname
\fi
#2}}
\providecommand{\BIBdecl}{\relax}
\BIBdecl

\bibitem{D06}
D.~Donoho, ``Compressed sensing,'' \emph{IEEE Trans. Inform. Theory}, vol. 52
  (4), pp. 1289--1306, April 2006.

\bibitem{CRTFeb06}
E.~Cand\`{e}s, J.~Romberg, and T.~Tao, ``Robust uncertainty principles: Exact
  signal reconstruction from highly incomplete frequency information,''
  \emph{IEEE Trans. Inform. Theory}, pp. 489--509, February 2006.

\bibitem{XH07}
W.~Xu and B.~Hassibi, ``Efficient compressive sensing with deterministic
  guarantees using expander graphs,'' in \emph{Proc. Information Theory
  Workshop (ITW)}, September 2007, pp. 414--419.

\bibitem{AJWAK10}
A.~Abdelkefi, Y.~Jiang, W.~Wang, and A.~Kvittem, ``Robust traffic anomaly
  detection with principal component pursuit,'' in \emph{Proc. 6th ACM Int.
  Conf. on emerging Networking EXperiments and Technologies (CoNEXT) Student
  Workshop}, 2010.

\bibitem{LMP08}
Y.~Lu, A.~Montanari, and B.~Prabhakar, ``Counter braids: Asymptotic optimality
  of the message passing decoding algorithm,'' in \emph{46th Annual Allerton
  Conference on Communication, Control, and Computing}, September 2008, pp. 209
  -- 216.

\bibitem{LMPDK08}
Y.~Lu, A.~Montanari, B.~Prabhakar, S.~Dharmapurikar, and A.~Kabbani, ``Counter
  braids: A novel counter architecture for per-flow measurement,'' in
  \emph{Proc. Int. Conf. on Measurement and Modeling of Computer Sys. ACM
  SIGMETRICS}, June 2008, pp. 121--132.

\bibitem{CT05}
E.~Cand\`{e}s and T.~Tao, ``Decoding by linear programming,'' \emph{IEEE Trans.
  Inform. Theory}, vol.~51, pp. 4203--4215, Dec. 2005.

\bibitem{MYLHH10}
J.~Meng, W.~Yin, H.~Li, E.~Houssain, and Z.~Han, ``Collaborative spectrum
  sensing from sparse observations using matrix completion for cognitive radio
  networks,'' in \emph{Proc. IEEE Int. Conf. Acoustics, Speech, and Signal
  Proc. (ICASSP)}, March 2010, pp. 3114 -- 3117.

\bibitem{HB11}
\BIBentryALTinterwordspacing
R.~Heckel and H.~Bolcskei, ``Compressive identification of linear operators.''
  [Online]. Available: \url{http://arxiv.org/abs/1105.5215}
\BIBentrySTDinterwordspacing

\bibitem{ZP07J}
F.~Zhang and H.~D. Pfister, ``Analysis of verification-based decoding on the
  q-ary symmetric channel for large q,'' \emph{IEEE Trans. Inform. Theory},
  vol. 57 (10), pp. 6754--6770, 2011.

\bibitem{ZP09}
\BIBentryALTinterwordspacing
------, ``On the iterative decoding of high rate {LDPC} codes with applications
  in compressed sensing.'' [Online]. Available:
  \url{http://arxiv.org/abs/0903.2232}
\BIBentrySTDinterwordspacing

\bibitem{EHBL11}
\BIBentryALTinterwordspacing
Y.~Eftekhari, A.~Heidarzadeh, A.~Banihashemi, and I.~Lambadaris, ``An efficient
  approach toward the asymptotic analysis of node-based verification-based
  algorithms in compressed sensing,'' \emph{Submitted to IEEE Trans. Inform.
  Theory}, April 2011. [Online]. Available:
  \url{http://arxiv.org/abs/1104.0224}
\BIBentrySTDinterwordspacing

\bibitem{EHBLISIT11}
------, ``An efficient approach toward the asymptotic analysis of node-based
  verification-based algorithms in compressed sensing,'' in \emph{Proc. IEEE
  Int. Symp. Inform. Theory (ISIT)}, 2011.

\bibitem{LM05}
M.~Luby and M.~Mitzenmacher, ``Verification-based decoding for packet-based
  low-density parity-check codes,'' \emph{IEEE Trans. Inform. Theory}, vol. 51
  (1), pp. 120--127, January 2005.

\bibitem{ZP07}
F.~Zhang and H.~D. Pfister, ``List-message passing achieves capacity on the
  q-ary symmetric channel for large q,'' in \emph{IEEE Global Telecom. Conf.
  (GLOBECOM)}, November 2007, pp. 283--287.

\bibitem{BGIKS08}
R.~Berinde, A.~Gilbert, P.~Indyk, and K.~Strauss, ``Combining geometry and
  combinatorics: A unified approach to sparse signal recovery,'' in \emph{46th
  Annual Allerton Conference on Communication, Control, and Computing},
  September 2008, pp. 798--805.

\bibitem{SBB206}
S.~Sarvotham, D.~Baron, and R.~Baraniuk, ``Sudocodes - fast measurement and
  reconstruction of sparse signals,'' in \emph{Proc. IEEE Int. Symp.
  Information Theory (ISIT)}, July 2006, pp. 2804--2808.

\bibitem{ZP08}
F.~Zhang and H.~D. Pfister, ``Compressed sensing and linear codes over real
  numbers,'' in \emph{Proc. Information Theory and Applications Workshop},
  February 2008, pp. 558--561.

\end{thebibliography}
\end{document}